\documentclass[onecolumn,floatfix,pra,superscriptaddress,nofootinbib]{revtex4}
\usepackage{color}
\usepackage{graphicx}

\usepackage{enumerate}

\usepackage{amsmath,amssymb,amsfonts}
\usepackage{bm}

\newcommand{\be}{\begin{equation}}
\newcommand{\ee}{\end{equation}}

\def\ba{\begin{eqnarray}}
\def\ea{\end{eqnarray}}

\begin{document}
	

\title{Quantum surface diffusion in Bohmian Mechanics}
\author{S. Miret-Art\'es}  
\email{s.miret@iff.csic.es}

\address{Instituto de F\'isica Fundamental, Consejo Superior de
Investigaciones Cient\'ificas, Serrano 123, 28006 Madrid, Spain}
\begin{abstract}
Surface diffusion of small adsorbates is analyzed in terms of the so-called intermediate scattering function
and dynamic structure factor, observables in experiments using the well-known quasielastic Helium atom 
scattering and Helium spin echo techniques. The linear theory applied is an extension of the neutron scattering due
to van Hove and considers the time evolution of the position of the adsorbates in the surface. This approach
allows us to use a stochastic trajectory description following the classical, quantum and Bohmian 
frameworks. Three different regimes of motion are clearly identified in the diffusion process: ballistic, Brownian 
and intermediate which are well characterized, for the first two regimes, through the mean square 
displacements and Einstein relation for the diffusion constant. The Langevin formalism is used by 
considering Ohmic friction, moderate surface temperatures and small coverages. In the Bohmian framework, analyzed here,
the starting point is the so-called Schr\"odinger-Langevin equation which is a nonlinear,
logarithmic differential equation. By assuming a Gaussian function for the probability density, the 
corresponding quantum stochastic trajectories are given by a dressing scheme consisting of a classical 
stochastic trajectory followed  by the center of the Gaussian wave packet, and issued from solving the Langevin equation 
(particle property), plus the time evolution of its width governed by the damped Pinney differential equation 
(wave property). The Bohmian velocity autocorrelation function is the same as the classical one when the initial spread 
rate is assumed to be zero. If not, in the diffusion regime, the Brownian-Bohmian motion shows
a weak anomalous diffusion.  

\end{abstract}

\maketitle


	



\section{Introduction}

Surface diffusion is one of the most elementary dynamical process occurring on surfaces and a 
preliminary step to more complex surface phenomena. It is a very active field of surface science from 
fundamental as well as technological (catalysis, crystal growth, energy storage, etc.) points of view. 
Typically, this diffusion process is analyzed as in spectroscopic 
experiments where a probe particle is interacting perturbatively with a given system at thermal equilibrium 
with a reservoir (or thermal bath) and measuring its response. According to van Hove's theory for neutron
scattering by crystal and liquids \cite{vanhove,vineyard,lovesey}, the nature of particles (photons, 
neutrons, electrons or atoms) probing  systems of moving and interacting particles (adsorbates)  is largely 
irrelevant when the Born approximation is assumed, reducing this scattering event to a typical statistical 
mechanics problem. The corresponding linear response is then determined by 
the spectrum of the spontaneous fluctuations of the reservoir as established by the very well-known
fluctuation-dissipation theorem \cite{kubo}. Information provided by the experiment together with a 
theoretical support or theory behind can allow us to better understand the dynamics as well as extract 
valuable information for molecular interactions (adsorbate-substrate and adsorbate-adsorbate interactions) 
within the general framework of stochastic processes. 
A very large amount of information about the diffusion process in surfaces has been gathered along
the last twenty eight years from the well-known review paper by Gomer \cite{gomer}.        
For fast diffusion motions, we are going to focus on He atoms as nondestructive probe particles used 
in two types of experiments,
quasielastic He atom scattering (QHAS) \cite{peter} and He spin echo spectroscopy (HeSE) \cite{bill}. 
These time of flight techniques are sensitive to surface processes on the length and time scales on which 
single atoms diffusion occurs (length scales between around 10$^{-10}$ up to 10$^{-8}$ meters and 
time scales going from around 10$^{-12}$ up to 10$^{-8}$ seconds). Time of flight spectra are usually 
converted to energy transfer scale allowing a frequency analysis of the
surface phonons as well as slow motions of the adsorbates. Angular (around 0.3$^0$) as well as velocity 
(around 1 $\%$) resolutions are very small covering a large dynamical range in intensity; much better for 
the HeSE technique. Typical He velocities are less than $3 \times 10^{3}$ m/sec. The practical limit of these
techniques lies in the velocity spread in the beam but, with the spin-echo method, one measures velocity 
changes of individual atoms rather than the velocity change with respect to the mean incident velocity. 
The major challenges facing these techniques are to analyze and extract valuable information from the observed 
line shapes as well as time behavior.

Van Hove's theory of neutrons was generalized to atom surface scattering within the transition matrix 
formalism \cite{dick} and  the Chudley-Elliott aproximation \cite{frenken,chudley}. 
In surface diffusion problems, most of work is based on the Langevin equation
formalism which is widely used when dealing with stochastic processes, as the diffusion one. Thanks to
Caldeira and Leggett, this formalism can be derived from a Hamiltonian which is split into three parts  
describing the dynamics of the system, the thermal bath or reservoir and their mutual interaction
\cite{caldeira,weiss}. The surface is usually considered to be corrugated and, at a given temperature, is
replaced by an infinite number of harmonic oscillators, mimicking the phonon dynamics as well as
the mechanism of dissipation. An Ohmic friction is typically assumed and the damping mechanism is
mainly due to acoustic phonons. For barriers greater than $3 k_B T$ ($k_B$ is the Boltzmann factor and $T$ 
the surface temperature), the diffusion process is activated and the instantaneous jump picture works quite well.
Activation barrier heights are extracted from an Arrhenius plot of the diffusion coefficient. Large discrepancies
are obtained when comparing the experimental or theoretical results to the classical transtiton state theory
\cite{hofmann} due to the existence of long jumps at high surface temperatures, multiple jumps where
the Chudley-Elliott model does not apply. A quantum and classical Kramer's theory was developed to
overcome such discrepancies \cite{melnikov,eli,salva1,salva2,salva3}, leading to analytic expressions for 
diffusion coefficients, escape rates and hopping distributions within the Langevin formalism.   

Whenever the diffusing atoms are light such as hydrogen or deuterium, quantum effects are present. It
is known that quantum diffusion coefficients can be smaller or greater than the classical ones \cite{eli}.
For example, if the substrate is Pt(111), Arrhenius plots of the diffusion constant and overall hoping rate show 
clearly a region where deviations from the linearity are observed, which is characteristic of the 
classical transition state theory (TST) \cite{bill2}. This deviation starts occuring at low temperatures 
(below 90 K) and the theory of dissipative tunneling \cite{wolynes}, based on the quantum TST, is sufficient 
to be applied. The flattening of the Arrhenius plot at the crossover temperature is however not observed 
which is a feature of deep tunneling \cite{weiss}. In this regime, Grabert and Weiss accounted for quantum 
diffusion in periodic potentials \cite{grabert1,grabert2} by using the so-called bounce technique together with
the Chudley-Elliott model, leading to analyticl expressions for  transition rates and diffusion constants in 
an incoherent tunneling regime. This theoretical framework was successfully applied to this diffusion problem  
\cite{salva4} for low coverages. In any case, as far as we know, this interesting and particular quantum
dynamics has not been analyzed in the Langeving formalism, that is, by using quantum stochastic trajectories.

In this work, a natural theoretical approach considering quantum trajectories is analyzed within Bohmian mechanics 
which is being more and more applied to conservative and open problems \cite{holland,salva5,salva6,salva7}.  
Recently, an extension to open quantum systems  (see, stochastic processes), within 
the nonlinear, logarithmic Schr\"odinger-Langevin (SL) equation framework derived by Kostin
\cite{kostin}, has been proposed under the presence or not of a continuous measurement 
\cite{salva7,antonio1} and for nonlinear dissipation  \cite{pedro}. The resulting quantum stochastic
trajectories have been applied to simple systems such as the damped free particle, linear potential, 
and harmonic oscillator\cite{vahid1} and dissipative quantum tunnelling through an inverted parabolic barrier
under the presence of an electric field \cite{vahid2} when analysing the classical-quantum transition of 
trajectories in the gradual decoherence process. These works introduced the so-called scaled trajectories
having as a particular case the Bohmian ones. By assuming a time-dependent Gaussian ansatz for the 
probability density, theses scaled trajectories are written as a sum of  a classical trajectory followed by the center 
of the Gaussian wave packet (a particle property) plus a term containing the time evolution of its width (a wave property) within 
of what has been called {\it dressing scheme} \cite{salva7}.  

The organization of this work is as follows. In Section II, the general theory for neutron scattering due to 
van Hove is briefly reviewed to better understand the extension to atom scattering. Two main observable 
functions the so-called dynamic structure factor and intermediate scattering function are introduced and written in terms
of adsorbate trajectories. These trajectories are briefly presented and discussed within the general Langevin formalism 
starting from the so-called Caldeira-Leggett Hamiltonian in the classical and quantum frameworks, being the adsorbate coverage  introduced 
by a collisional friction. In this way, the Bohmian 
framework developed afterwards in terms of the Schr\"odinger-Langevin equation is easier to follow.  In Section III,  three main
different  regimes in the diffusion process are well characterized and analyzed in the classical and quantum domains: the ballistic, 
Brownian (or diffusion) and intermediate regimes. For each case, the corresponding trajectories are analyzed in terms of the mean square 
displacements and velocity autocorrelation functions leading to analytical expressions for the observable lines shapes. In the 
second regime, the Brownian-Bohmian motion shows a weak anomalous diffusive behavior.

\section{General theory}

\subsection{Observables}

In 1954, van Hove \cite{vanhove} established the differential cross section of the scattering 
of slow neutrons by a system of interacting particles in terms of the generalized pair distribution
function, the so-called $G({\bf r},t)$ function of van Hove (with ${\bf r}$ being a position vector and
$t$ a time interval). This $G$ function is a natural extension of the standard pair distribution function 
$g({\bf r})$ well known, for example, in liquids with $G({\bf r},0) = g({\bf r})$. Moreover, $G$ describes 
the correlation between a particle in position ${\bf r}+ {\bf r}'$ at $t+t'$ and a particle in position 
${\bf r}'$ at time $t'$ . In the Born approximation or first order perturbation theory, the scattering problem 
is reduced essentially to a problem in statistical mechanics \cite{vanhove,vineyard,lovesey} where the nature of 
the scattered particles (neutrons, light, atoms, etc.) and details of the interaction potential are irrelevant. 
In this formalism, the linear response of the system implies that it is determined entirely by the 
properties exhibited by the system in the absence of probe particles. This differential cross section 
can also be written in terms of the independent variables associated with the momentum transfer,
$\hbar {\bf k}$, and energy transfer $\hbar \omega$ as
\be
\frac{d^2 \mathcal{R} ({\bf k}, \omega)}{d\Omega d\omega} \propto
S({\bf k}, \omega) = (2 \pi)^{-1} N \int e^{i ({\bf k}. {\bf r} - \omega t)} G({\bf r},t) d{\bf r} dt
\label{dsf}
\ee
providing the probability that the probe particles scattered from the diffusing system 
reach a certain solid angle $\Omega$ in an interval of outgoing energy $\hbar \omega$.
The response function or line shape $S({\bf k}, \omega)$ is also termed the scattering law or dynamic 
structure factor (DSF) where $N$ is introduced for convenience and represents the number of interacting 
particles in the system under study. The spatial Fourier transform of the $G$-function 
\be
I({\bf k}, t) = \int e^{i {\bf k}. {\bf r}} G({\bf r},t) d{\bf r} 
\label{isf}
\ee
is called intermediate scattering function (ISF) and therefore $S$ and $I$ are related by the inverse Fourier 
transform in time. These functions are easily showed to be expressed in terms of the density-density 
correlation function where the particle density operator is defined as
\be
\rho ({\bf k}, t) = \sum_j^N \delta({\bf r}-{\bf r}_j(t)) 
\label{rho}
\ee

In this work, we are going to describe on the QHAS technique 
probing the dynamics of adsorbates or adparticles on surfaces \cite{hofmann}. With this technique, at 
thermal energies, time--of--flight measurements of the probe particles are converted to energy transfer 
spectra given by the dynamics structure factor.  In this scattering, He atoms presents an energy exchange 
$\hbar \omega =E_{final} - E_{initial}$ and a parallel (to the surface) momentum transfer 
${\bf K} = {\bf K}_{final} - {\bf K}_{initial}$ (it is standard to express variables projected on the surface 
as capital letters for position ${\bf R}=(x,y)$ and parallel momentum ${\bf K}$).
The prominent peak around the zero energy transfer, the so-called quasi--elastic peak (Q--peak), 
provides direct information of adsorbate diffusion. Additional weaker peaks at low 
energy transfers around the Q--peak are also observed and  attributed to the parallel 
frustrated translational motion of some adsorbates (the so-called T--mode) and to 
surface phonons excitations. Long distance and time correlations are extracted from the scattering 
law when considering small values of ${\bf K}$ and $\hbar \omega$, respectively.
The  nature of the  adsorbate--substrate and adsorbate--adsorbate interactions can also be known from
the scattering law. In this context, the dynamic structure factor is usually expressed as 
\be
S({\bf K},\omega) = (2 \pi)^{-1} N
\int e^{-i\omega t} \ \! I({\bf K},t) \ \! dt ,
\label{dsf1}
\ee
with
\be
I({\bf K},t) \equiv \frac{1}{N} 
\langle \sum_{j,j'}^N e^{-i {\bf K} \cdot
	{\bf R}_j (0)} e^{i  {\bf K} \cdot {\bf R}_{j'}(t)} \rangle
\label{isf1}
\ee
where the brackets denote an ensemble average and ${\bf R}_j (t)$ the position vector of the $j$ adparticle 
at time $t$ on the surface. This intermediate scattering function is precisely what is
directly measured from the HeSE technique \cite{bill} which is quite similar to the well 
known neutron spin echo one.  

At this point, it is important to stress the main difference between neutron and Helium scattering.
The $G$-function can naturally be split into a part describing the correlations between the same particle,
$G_s$, and distinct particles, $G_d$, where the crossing terms are taken into account.  Thus, the full 
pair correlation function can then be expressed as
\begin{equation}
G({\bf R},t) = G_s({\bf R},t) + G_d({\bf R},t) .
\label{gtotal}
\end{equation}
According to its definition, $G_s({\bf R},0) = \delta({\bf R})$ which the Dirac delta function gives the 
presence of the particle at that position and $G_d({\bf R},0) = g({\bf R})$.  
At low adparticle concentrations (coverage, $\theta \ll 1$), when interactions among adsorbates can be 
neglected because they are far apart from each other, the main contribution to (\ref{gtotal}) is $G_s$
(particle--particle correlations are negligible and $G_d \approx 0$).
On the contrary, at high coverages, $G_d$ is expected to have a significant contribution to (\ref{gtotal}).
As a result of this splitting, the intermediate scattering function can also be expressed as a sum of distinct 
($I_d$) and self ($I_s$) functions.  Following neutron scattering language, the
corresponding Fourier transforms of $I$ and $I_s$ give the so-called coherent scattering law, 
$S({\bf K},\omega)$ and incoherent scattering law $S_s ({\bf K},\omega)$, respectively.
In QHAS and HeSE experiments, only coherent scattering is observed.

After Eq. (\ref{isf1}), the ISF contains information about the dynamics of the 
adsorbates through ${\bf R}_j (t)$. This dynamics is open since the surface can be seen as a reservoir or
thermal bath at a given temperature, leading to dissipation and stochasticity within a classical or quantum 
framework. In the following, we are going to focus on the nature of the adsorbate-adsorbate 
and adsorbate-substrate interactions. In any case, a proper comparison between the experimental and 
theoretical observables (issued from any theoretical method) has to be carried out through a convolution 
integral which takes into account the response of the apparatus which is usually assumed a Gaussian function.

\subsection{Classical stochastic trajectories}

For heavy adsorbates, the time-dependent position vectors can be obtained from classical stochastic trajectories.
As mentioned above, if the coverage is very small, the adsorbate-adsorbate interaction is negligible and the
dynamics can be well described only by the self part of the $G$-function, $G_s$. The main interaction is then
the adsorbate-substrate interaction as well as the thermal fluctuations of the surface through a random force or
noise. In the literature, the standard Hamiltonian used is that proposed by Magalinskij \cite{magalinski} and
Caldeira and Leggett \cite{caldeira}  written in this context as \cite{salva3,salva8}
\begin{eqnarray}
H & = & \frac{p_x^2}{2m} + \frac{p_y^2}{2m} + V(x,y)  + \frac{1}{2} \sum_{j=1}^{N} 
\left[ \frac{p_{x_j}^2}{m_j}+ m_j \omega^2_{x_j}
\left(  x_j - \frac{c_{x_j}}
{m_j \omega^2_{x_j}} \ \! x \right)^2  \right]
\nonumber \\
& & + \frac{1}{2} \sum_{j=1}^{N} 
\left[ \frac{p_{y_j}^2}{m_j}+ m_j \omega^2_{y_j}
\left(  y_j - \frac{c_{y_j}}
{m_j \omega^2_{y_j}} \ \! y \right)^2  \right]  ,
\label{hcl}
\end{eqnarray}
where $(p_x,p_y)$ and $(x,y)$ are the adparticle momenta and positions with mass $m$, $(p_{x_i},x_i)$ 
and $(p_{y_i},y_i)$ with $i = 1, \cdots , N$ are the momenta and positions of the bath oscillators 
(phonons) for each degree of freedom, with mass and frequency given by $m_i$ and $\omega_i$, respectively.
Phonons with polarization along the $z$--direction are not considered. The adsorbate-substrate interaction
$V(x,y)$ is  a periodic function describing the surface corrugation at zero temperature. The Hamiltonian 
(\ref{hcl}) is not translational invariance since the term coupling the parallel motions to the phonon bath in 
both directions is not periodic but linear \cite{jeremy}. However, this Hamiltonian is still used because it 
leads to the correct generalized Langevin equation once the bath degrees of freedom are eliminated
\begin{subequations}
		\ba
		m \ddot{x}(t) & + & \displaystyle m \int_0^t \gamma_x (t-t') \
		\dot{x}(t') \ dt' + \frac{\partial V(x,y)}{\partial x} = \xi_x (t) , \\
		m \ddot{y}(t) & + & \displaystyle m \int_0^t \gamma_y (t-t') \
		\dot{y}(t') \ dt' + \frac{\partial V(x,y)}{\partial y} = \xi_y (t) ,
		\ea
		\label{gle}
\end{subequations}
where the friction coefficients are defined through the cosine Fourier transform of the spectral densities,
\be
\gamma_i (t) = \frac{2}{\pi m} \int_0^{\infty}
\frac{J_i (\omega)}{\omega}\ \cos \omega t\ d \omega,
\label{fric}
\ee
with $i=x,y$ and  
\begin{equation}
J_i(\omega) = \frac{\pi}{2}
\sum_{j=1}^N \frac{c_{i_j}^2}{m_j \omega_{i_j}^2}
\left[ \delta (\omega - \omega_{i_j}) \right] .
\label{sd}
\end{equation}
The nonhomogeneity of (\ref{gle}) represents a fluctuating or random force $\xi$ for each 
degree of freedom which depends  on the initial position of the system and initial positions and momenta 
of the oscillators of each bath  according to \cite{weiss}
\be
\xi_x(t) = - \sum_j c_{x_j} \left\{ \left[x_j(0)
+ \frac{c_{x_j}(0)}{m_j \omega_{x_j}^2} \ x(0) \right] \cos (\omega_{x_j} t)
+ \frac{p_{x_j}(0)}{m_j \omega_{x_j}}\ \sin (\omega_{x_j} t) \right\} .
\label{spb-7a}
\ee
and 
\be
\xi_y(t) = - \sum_j c_{y_j} \left\{ \left[y_j(0)
+ \frac{c_{y_j}(0)}{m_j \omega_{y_j}^2} \ y(0) \right] \cos (\omega_{y_j} t)
+ \frac{p_{y_j}(0)}{m_j \omega_{y_j}}\ \sin (\omega_{y_j} t) \right\} .
\label{spb-7b}
\ee
If Ohmic friction is assumed, $\gamma_i (t) = 2 \gamma_i \delta (t)$, where $\gamma_i$ is a constant and 
$\delta (t)$ is Dirac's $\delta$--function. Eqs. (\ref{gle}) then reduce to two coupled standard Langevin
equations (the	$\delta$--function counts only one half when the integration is
carried out from zero to infinity) 
\begin{subequations}
	\ba
	m \ddot{x}(t) & + &  m \gamma_x  \
	\dot{x}(t)  + \frac{\partial V(x,y)}{\partial x} = \xi_x (t) , \\
	m \ddot{y}(t) & + &  m  \gamma_y  \
	\dot{y}(t)  + \frac{\partial V(x,y)}{\partial y} = \xi_y (t)   .
	\ea
	\label{le}
\end{subequations}
within the Markov approximation. The properties of noise are: (i) $\left\langle  \xi_i (t) \right\rangle =0$ 
(zero mean) and (ii) $\left\langle  \xi_i (0)  \xi_i(t) \right\rangle = m k_B T \gamma_i$, with $i=x,y$. 
The corresponding classical stochastic trajectories are given by ${\bf R} (t) = (x(t),y(t))$.
When a flat surface is considered, $V(x,y)=0$ and the standard Brownian motion takes place. 

At higher coverages, adsorbate--adsorbate interactions can no longer be neglected and typically pairwise
interaction potentials  are usually introduced in Langevin molecular dynamics simulations \cite{john}.
These simulations always result in a relatively high computational cost due to the time spent by the codes 
in the evaluation of the forces among particles. This problem is even worse when working with 
long--range interactions, since {\it a priori} they imply that one should consider a relatively
large number of particles to numerical convergency. An alternative approach 
is to consider a purely stochastic description for these interactions \cite{ruth1,ruth2,ruth3} through what
is called the interacting single adsorbate  (ISA) approximation  in a two-bath model.  The motion of a 
single adsorbate is then modelled by a series of random pulses within a Markovian regime (i.e., pulses of
relatively short duration in comparison with the system relaxation and acting during a long period of time).
These pulses simulate the collisions among adsorbates and are described by means of a white shot noise. 
In this way, a typical molecular dynamical simulation problem involving $N$ adsorbates is substituted 
by the dynamics of a single adsorbate where the action of the remaining $N-1$ adparticles is replaced by 
a random force given by the white shot noise. The surface coverage is related to a collisional friction
providing the average number of collisions per unit time, $\gamma_c$. The probability of observing a 
given number of  collisions, after an elapsed  time, follows closely a Poisson distribution. The adsorbate 
is then subject to two uncorrelated white noises, one coming from the substrate and the other one from the
surrounding adsorbates. Thus, the total friction coefficient $\eta$ in the ISA approximation is a sum of two
friction coefficients, $\eta = \gamma + \gamma_c$ and the total noise is given by
$\xi = \xi_G + \xi_S$ (where $G$ stands for  Gaussian and $S$ for shot) for each degree 
of freedom of the surface ($x,y$). In this way, differences between self and distinct time--dependent pair
correlation function do not exist but Eqs.~(\ref{dsf}) and ~(\ref{isf}) still hold. The ISF can now be rewritten as

\be
I({\bf K},t) \equiv
\langle e^{-i {\bf K} \cdot
	[{\bf R}(t) - {\bf R}(0)] } \rangle
= \langle e^{-i {\bf K} \cdot
	\int_0^t {\bf v} (t') \ \! dt'} \rangle        .
\label{isf-isa}
\ee
Within the so--called {\it Gaussian approximation} \cite{mcquarrie}, which is exact when the velocity
correlations at more than two different times are negligible, Eq. (\ref{isf-isa})  is expressed again as
a second order cumulant expansion in ${\bf K}$ 
\begin{equation}
I({\bf K},t) \approx
e^{- K^2 \int_0^t (t - t')
	\mathcal{C}_{{\bf K}}(t') dt'} .
\label{isf-isa2}
\end{equation}
with 
\begin{equation}
\mathcal{C}_{{\bf K}}(\tau) \equiv
\langle v_{{\bf K}}(0) \ \!
v_{{\bf K}}(\tau) \rangle =
\lim_{\mathcal{T}\to\infty} \frac{1}{\mathcal{T}}
\int_0^\mathcal{T} v_{{\bf K}}(t) \ \!
v_{{\bf K}}(t+\tau) \ \! dt
\label{vcorr}
\end{equation}
being the {\it velocity autocorrelation function} (VAF) projected onto the direction of the parallel momentum 
transfer. The velocity is considered to be a stationary stochastic process. This autocorrelation function 
decays with time, allowing us to define a characteristic time, the so-called correlation time, as
\begin{equation}
\tilde{\tau} \equiv \frac{1}{\langle v_0^2 \rangle}
\int_0^\infty \mathcal{C}_{{\bf K}}(\tau) \ \! d\tau ,
\label{tau}
\end{equation}
where $\sqrt{\langle v_0^2 \rangle} = \sqrt{k_B T/m}$ is the average thermal velocity in one dimension, 
along the direction given by ${\bf K}$, $m$, $T$ and $k_B$ being the adsorbate mass, surface 
temperature and Boltzmann constant, respectively.

The advantage of this approximation consists in providing a direct expression for the coherent scattering 
which is observed when He atoms are used as probe particles. The dynamical structure factor depends on
the VAF through the intermediate scattering function if the Gaussian
approximation is also assumed. Two extreme regimes are well characterized in this context, the ballistic
diffusion, at very small times ($\eta t << 1$), which is a frictionless motion and the diffusion regime, 
at very long times ($\eta  t >> 1$), when the thermal equilibrium is already reached and details of the
surface such as its corrugation is no longer important. Analytical expressions 
for line shapes in these two extreme regimes are easily derived due to have simple velocity autocorrelations
functions \cite{salva8}.

\subsection{Quantum Langevin equation}

When considering light adsorbates, quantum mechanics in the Heisenberg picture should be applied.
Quantum vector positions in Eqs. (\ref{isf1}) and (\ref{isf-isa}) are then seen as operators. 
At two different times, they do not commute. However, it is possible to factorize 
the  ISF in two factors due to the disentangling theorem according to 
$e^A e^B = e^{A+B} e^{[A,B]/2}$ which holds when the corresponding commutator is a $c$-number.
Thus, if $A=i {\bf K}. {\bf R}(0)$ and $B= - i  {\bf K} . {\bf R}(t)$ then \cite{salva8} 
\begin{equation}
I({\bf K},t) = I_1({\bf K},t)  I_2({\bf K},t) 
\label{i12}
\end{equation}
with the $I_2$-factor is given by Eq. (\ref{isf-isa2}). The $I_1$ factor can be readily obtained from the formal
solution of the corresponding Langevin equation (if the Ohmic friction is assumed) given by Eq. (\ref{le})
\begin{equation}
{\bf R}(t) = {\bf R}(0) + \frac{{\bf P}(0)}{m \eta} \Phi(\eta t)  + \frac{1}{m \eta} 
\int_0^t \Phi(\eta t-\eta t') \left[  {\bf F}({\bf R}(t')) + \delta {\bf F}_r (t') \right]   dt' 
\end{equation} 
where $\Phi(x) = 1- e^{-x}$, ${\bf F}({\bf R}) = - \nabla_{\bf R} V$, $\delta  {\bf F}_r (t)$ is the random 
force including the Gaussian and shot noises and ${\bf R}(0)$ and ${\bf P}(0)$ are the initial conditions for
the position and momentum, respectively. The commutator involved in $I_1$ is $ i \hbar$ since 
$[{\bf R}(0),{\bf P}(0)] = i \hbar$, $[{\bf R}(0),{\bf F}] = i \hbar \partial {\bf F} / \partial {\bf P}(0) = 0$ and
$[{\bf R}(0), \delta {\bf F}_r] = 0$  if the noise is assumed to be classical (moderate surface temperatures). Thus,
\begin{equation}
I_1({\bf K},t) = e^{i E_r  \Phi(\eta t)/ \hbar \eta}   
\label{i1}
\end{equation}
where $E_r = \hbar^2 {\bf K}^2 / 2 m$ is the adsorbate recoil energy. $I_1$ is a time
dependent phase factor which is less and less important when the adsorbate mass and the friction 
coefficient increase.

\subsection{Bohmian stochastic trajectories}

Once the classical and quantum Langevin approach have been briefly reviewed, the Bohmian formalism is easier 
implemented and understood. An alternative way to describe the quantum diffusion motion is through Bohmian (or quantum) stochastic
trajectories. For this goal, we start from the so-called SL o Kostin equation \cite{kostin}. 
In 1972, Kostin derived heuristically this equation  from the standard one-dimensional Langevin equation. In this context, 
from Eq. (\ref{le}), the corresponding nonlinear two-dimensional Schr\"odinger equation is written as  
\begin{equation}
i\hbar\ \frac{\partial \Psi({\bf R},t)}{\partial t} =
- \frac{\hbar^2}{2 m} \nabla^2 \Psi({\bf R},t) 
 +  \left [V({\bf R}) + V_r ({\bf R},t) + V_D ({\bf R},t) + G(t) \right]  \Psi({\bf R},t)   , 
\label{kostin}
\end{equation}
where the random potential is given by
\begin{equation}
V_r ({\bf R},t)= - {\bf R}.{\bf F}_r(t)   , 
\label{random}
\end{equation}
the damping potential  by
\be
V_D({\bf R},t)  =  - \frac{i \hbar \eta}{2} \ln \left( \frac{\Psi({\bf R},t)}{\Psi^*({\bf R},t)} \right)  
\label{damping}
\ee
and
\begin{equation}
G(t) =  \frac{i \hbar \eta}{2}\
\int \Psi^*({\bf R},t) \ln \left( \frac{\Psi({\bf R},t)}{\Psi^*({\bf R},t)} \right)  
\Psi({\bf R},t) d {\bf R}
\label{gt}
\end{equation}
is a time dependent function resulting from the average value of $V_D$ by integration with 
respect to the position variable. The norm of the wave function is conserved and the expectation value 
of the corresponding nonlinear Hamiltonian is, as usual, the sum of the kinetic and potential energies at 
any time. The SL equation does not fulfill the superposition principle. 

If the wave function is written in  polar form as  
\begin{equation}
\Psi ({\bf R},t) = \phi({\bf R},t) e^{i S({\bf R},t) / \hbar}
\label{polar}
\end{equation}
where $\phi({\bf R},t)$ and $S({\bf R},t)$ are real valued functions and  then is substituted into 
Equation (\ref{kostin}), the resulting Schr\"odinger-Langevin-Bohm (SLB) 
equation reads as \cite{salva7}
\begin{eqnarray} \label{kostin-bohmian} 
	i\hbar \left[ {\frac{{\partial \phi }}{{\partial t}} + \frac{i}{\hbar }\frac{{\partial S}}
		{{\partial t}}\phi } \right] & = & 
	- \frac{{{\hbar ^2}}}{{2m}}\left\{ \left[ \nabla ^2 \phi  
			- \frac{\phi }{\hbar ^2} \left(\nabla S \right)^2 \right] 
		+ \frac{i}{\hbar }\left[2 \nabla S \nabla \phi  + \phi \nabla ^2 S \right] \right\} \nonumber \\
	& + & \left [ V({\bf R}) + V_r({\bf R},t) +  \eta  ( S - \langle S \rangle)  \right] \phi.
\end{eqnarray}
\noindent 
Now, by writing the real and imaginary parts separately, we readily reach the continuity equation
\begin{equation}
	\frac{\partial \rho}{\partial t} + \nabla . (\rho {\bf v}) = 0
	\label{continuity}
\end{equation}
\noindent 
with $\rho = \phi^2$ and the velocity field defined by ${\bf v}= \nabla S / m$, and the 
quantum dissipative Hamilton-Jacobi equation given by 
\begin{equation}
	\frac{\partial {\bf v}}{\partial t} + {\bf v} . \nabla {\bf v} + \eta {\bf v} = - \frac{1}{m} 
	\nabla (V + V_r + Q),
	\label{hj}
\end{equation}
\noindent 
$Q$ being the quantum potential defined in terms of $\rho$ as follows 
\be
Q \equiv - \frac{\hbar^2}{2m}
\frac{\nabla^2 \rho^{1/2}}{\rho^{1/2}}
= \frac{\hbar^2}{4m}
\left[ \frac{1}{2} \left( \frac{\nabla \rho}{\rho} \right)^2
- \frac{\nabla^2 \rho}{\rho} \right] .
\label{q}
\ee

If a Gaussian ansatz is assumed for the probability density
\be
\rho ({\bf R},t)= \frac{1}{\sqrt{2 \pi \delta(t)^2}} e^{ - ({\bf R}-{\bf q}(t))^2 / 2 \delta(t)^2}
\label{rho}
\ee 
where $\delta(t)$ and ${\bf q}(t)$ are the width and the center of the wave packet, respectively.
From Eq. (\ref{continuity}), the velocity field turns out to be 
\be 
{\bf v}({\bf R},t) = \frac{\dot{\delta}(t)}{\delta(t)} ({\bf R} - {\bf q}(t)) + \dot{{\bf q}}(t)
 \label{vq}
\ee
where the dot on the variable means time derivation. The time integration
of this velocity field is straightforward leading to the equation for the Bohmian stochastic trajectories
\be
{\bf R}(t) = {\bf q}(t) +  ({\bf R}(0) - {\bf q}(0)) \frac{\delta(t)}{\delta(0)}   .
\label{qtray}
\ee
Eq. (\ref{qtray}) is given by a sum of a particle property through a classical
trajectory followed by the center of the wave packet plus a wave property involving the time evolution of its width. This scheme is 
known as {\it dressing scheme} \cite{salva7} which is issued only from the continuity equation (\ref{continuity}).
Now, substitution of Eq. (\ref{vq}) into Eq. (\ref{hj}) and after lengthy but straightforward calculations,
we reach  
\begin{eqnarray} \label{xterms} 
\left[ \frac{\ddot {\delta} (t)   +  {\eta} \dot \delta (t)}{\delta (t) } + \omega^2 - \frac{\hbar ^2}{4{m^2}{\delta ^4}(t)} \right ]
\left( {\bf R} - {\bf q}(t)	\right)^1
 +  \left [  \ddot {{\bf q}}(t) +  \eta  \dot{{\bf q}} (t) + \frac{1}{m}\nabla (V + V_r)|_{{\bf q}} \right ]
\left( {\bf R} - {\bf q} (t)  \right)^0 = 0,  \nonumber \\
\end{eqnarray}
leading to the standard Langevin equation for the center of the Gaussian wave packet when a Taylor expansion 
of the interaction potential around ${\bf q}$ up to second order is developed
\begin{equation}\label{leq}
\ddot {{\bf q}}(t) +  \eta  \dot{{\bf q}} (t) + \frac{1}{m}\nabla (V + V_r)|_{{\bf q}}= 0
\end{equation}
and the so-called dissipative or damped  Pinney equation for its width  
\begin{equation}\label{pinney}
\ddot {\delta} (t)   +  {\eta} \dot{\delta} (t) + \omega^2 \delta - \frac{\hbar ^2}{4{m^2}{\delta ^3}(t)} = 0   
\end{equation}
with $\omega = V''|_{\bf q} / m$.
The solution of this nonlinear differential equation was given by Pinney for the conservative case 
\cite{pinney} ($\eta = 0$) when $\hbar$ is replaced by an arbitrary constant. 

Zander et al \cite{zander} have also used the same ansatz to solve the Kostin equation under the presence of a 
continuous measurement. This procedure can also be seen as the "wave packet approximation" due to Gutzwiller \cite{gutzwiller}
where it is supposed that within the spatial range where the wave function is appreciably different from zero, the interaction potential
$V$ changes slowly enough so that it can be approximated to second order.

The commutation rule for the positions at different times does not work in this context. Moreover, the  
ISF given by Eq. (\ref{isf1}) can be replaced by Eq. (\ref{isf-isa}) within the ISA approximation and 
Eq. (\ref{isf-isa2}) when assuming the Gaussian approximation. Within this approximation,  the VAF
is the key function to be known or evaluated. The velocity of the quantum 
stochastic trajectories (\ref{qtray}) is readily obtained to be
\be \label{vq2}
{\bf v}({\bf R},t) = \frac{\dot{\delta }(t)}{\delta(0)} ({\bf R}(0) - {\bf q}(0)) + \dot{{\bf q}}(t)   
\ee
and the VAF along the ${\bf K}$ direction is then 
\begin{equation} \label{vcor}
\mathcal{C}_{{\bf K}}(\tau) \equiv
\langle v_{{\bf K}}(0) \ \!
v_{{\bf K}}(\tau) \rangle =
\langle \dot {q}(0) \dot {q}(\tau) \rangle_{{\bf K}} + \langle (R(0)- q(0))^2  \rangle_{{\bf K}}  \frac{\dot {\delta}(0)}{\delta^2(0)} \dot{\delta}(t)
\end{equation}
where cross correlations are zero due to the statistical independence.  It should be noticed that if the initial 
spread rate is assumed to be zero, $\dot{\delta} (0)=0$, the VAF behaves as in the classical regime.  
The quantum stochastic dynamics involved in the surface diffusion process within the Bohmian  framework 
and with the Gaussian ansatz is thus reduced to solve Eqs. (\ref{leq}) and (\ref{pinney}).

\section{Applications}

In surface diffusion, three different regimes of motion can be clearly distinguished. First, at very short times, $\eta t <<1$,
the motion is frictionless giving place to the so-called {\it ballistic regime}. Second, at very long times,
$\eta t >> 1$, the thermodynamical equilibrium has already been reached and we speak about the 
{\it Brownian or diffusion regime} and where the interaction with the surface is no longer relevant. 
And, finally, we have the  {\it intermediate regime} where the thermodynamical
equilibrium is still far to be reached. We pass now to analyze these three regimes within the classical,
quantum and Bohmian frameworks for comparison and provide analytical expressions (if possible) of the lines shapes within 
the Gaussian and ISA approximations. 

\subsection{The ballistic regime}

Due to the frictionless motion taking place at very short times (less than the mean free time), 
the corrugation of the surface plays no role in the surface dynamics. In the classical framework, the VAF
is expected to be constant with time and given by the thermal velocity along the ${{\bf K}}$ direction according to
\begin{equation}\label{vaf-br}
\mathcal{C}_{{\bf K}}(\tau) = \langle v_{{\bf K}}^2(0)  \rangle = \frac{k_B T}{m}    .
\end{equation}
From Eq. (\ref{isf-isa2}), we have
\begin{equation}
I({\bf K},t) \propto e^{-  K^2  \langle v_{{\bf K}}^2(0)  \rangle t^2/2}
\label{isf-isa-br}
\end{equation}
and from Eq. (\ref{dsf1})
\be
S({\bf K},\omega) \propto \frac{1}{|{\bf K}|  \sqrt{ \langle v_{{\bf K}}^2(0) \rangle}} e^{- \omega^2 / 2 K^2  \langle v_{{\bf K}}^2(0)  \rangle} ,
\label{dsf1-br}
\ee
which are the Gaussian behaviors predicted for both observables, the ISF and DSF or line shape.
This regime has been observed for a two dimensional free gas of Xe atoms on Pt(111) \cite{andrew,ruth4}. 
Thus, for times much shorter than the mean collision time, the
adsorbate displays a free motion showing a dynamical coherence since no memory lost of its velocity takes
place. Furthermore, the full width at half maximum (FWHM) of the line shape is linearly dependent on the
wave vector transfer, $\Gamma \propto   \sqrt{\langle v_{{\bf K}}^2(0) \rangle}   |{\bf K}| $.
In this ballistic regime, the mean square displacement (MSD) of the classical stochastic trajectories is known 
to be characterized by
\be
\langle |{\bf q}(t) - {\bf q}(0) |^2 \rangle \simeq \frac{k_B T}{m} t^2
\ee
showing a quadratic behavior with time.

In the quantum Langevin framework, the ISF is given by Eq. (\ref{i12}) together with 
Eqs. (\ref{isf-isa2}) and (\ref{i1}). As mentioned above, $I_1$ is a time dependent 
phase factor. In the limit of small times, $\Phi (\eta t) \sim \eta t$ and 
 \begin{equation}
 I_1({\bf K},t) = e^{i E_r  t/ \hbar }   .
 \label{i1-1}
 \end{equation}
The second factor $I_2$ is similar to the classical case and therefore
\begin{equation}
I({\bf K},t) \propto  e^{i E_r  t/ \hbar } e^{- K^2  \langle v_{{\bf K}}^2(0)  \rangle t^2/2}  .
\label{isf-isa-br-q}
\end{equation}
and 
\be
S({\bf K},\omega) \propto \frac{1}{ |{\bf K}|  \sqrt{ \langle v_{{\bf K}}^2(0) \rangle}} 
e^{- (\omega - E_r/\hbar)^2 / 2  K^2  \langle v_{{\bf K}}^2(0)  \rangle} .
\label{dsf1-br-q}
\ee
The Gaussian lineshape is thus shifted by the recoil energy whereas the FWHM is the same as before.

In the Bohmian framework, the starting point is Eq. (\ref{vcor}). At very short times, the adsorbate 
represented by a Gaussian function follows a free motion whose center is ruled by the simple 
differential equation  
\begin{equation}\label{leq-br}
\ddot {{\bf q}}(t)  = 0
\end{equation}
and its width is governed by the nondissipative Pinney equation   
\begin{equation}\label{pinney-br}
\ddot {\delta} (t)    - \frac{\hbar ^2}{4{m^2}{\delta ^3}(t)} = 0   .
\end{equation}
The solution of this nonlinear differential equation is \cite{antonio2}
\begin{equation}
\delta^2(t) = \delta^2(0) \left(1 + \frac{\dot {\delta}(0)}{\delta(0)} t \right)^2 +\frac{\hbar^2 t^2}
{4 m^2 \delta^2 (0)}
\end{equation}
which gives the standard time behavior for the width of a free Gaussian wavepacket when 
$\dot {\delta}(0) =0$,
\begin{equation}\label{delta1}
\delta (t) = \delta (0) \sqrt{1 + \left( \frac{\hbar t}{2 m \delta^2(0)} \right)^2}    .
\end{equation} 
In order to have the width contribution in Eq. (\ref{vcor}), we can assume, for example, that 
$\dot{\delta} (0) < \delta (0)$ (that is, the initial spreading rate is smaller than its initial width) leading,  in the so-called Fresnel or 
short time regime \cite{salva6}, to
\begin{equation}\label{delta2}
\delta (t) \approx \delta (0) +  \frac{\hbar^2 t^2}{8 m^2 \delta^3(0)} 
\end{equation} 
where the spreading increases quadratically with time. Thus, in the ballistic regime, and after Eq. (\ref{qtray}),
the Bohmian stochastic trajectories projected on ${\bf K}$ have the expression
\begin{equation}\label{qtray-br}
R(t) \approx v t - (R(0)-q(0)) \left( 1 + \frac{\hbar^2 t^2}{8 m^2 \delta^4(0)}\right)
\end{equation} 
where $v$ is the constant velocity of the adsorbate, $q(0)$ gives the initial condition for the center of the Gaussian 
wave packet and $R(0)$ is generated from the assumed initial Gaussian wave function.

On the other hand, the VAF along the ${\bf K}$  direction is expressed according to Eq. (\ref{vcor}) as
\begin{equation}
\mathcal{C}_{{\bf K}}(\tau) = \langle v_{{\bf K}}^2(0)  \rangle + 
\langle (R(0) - q(0))^2 \rangle_{{\bf K}}  \frac{\dot \delta (0)}{\delta(0)} 
\left( \delta (0) + \frac{\hbar^2}{8 m^2 \delta^4(0)} t^2\right)
\end{equation}
and  from Eq. (\ref{isf-isa2}), we have
\begin{equation}
I({\bf K},t) \propto e^{-  K^2  f( {\bf K},\dot{\delta}(0)) t^2/2}
\label{isf-isa-br}
\end{equation}
and from Eq. (\ref{dsf1})
\be
S({\bf K},\omega) \propto \frac{1}{ |{\bf K}|  \sqrt{f({\bf K},\dot{\delta}(0))}} 
e^{- \omega^2 / 2 K^2  f({\bf K},\dot{\delta}(0))} ,
\label{dsf1-br}
\ee
where
\begin{equation}
f({\bf K},\dot{\delta}(0)) = \langle v_{{\bf K}}^2(0)  \rangle + 
\langle (R(0) - q(0))^2 \rangle_{{\bf K}}  \dot \delta (0) 
\end{equation}
where only the first term in Eq. (\ref{delta2}) has been considered in order to keep constant the velocity
autocorrelation function which is the key point in the ballistic regime. The Gaussian functions thus 
obtained are different from those of the classical case except, as mentioned before, for the case where 
$\dot {\delta}(0) =0$. The commutation rule for the positions at different times is replaced, in this formalism, 
by the statistical choice of $R(0)$.

Finally, in this  regime, the MSD of the Bohmian or quantum stochastic trajectories is characterized by
\be
\langle |{\bf R}(t) |^2 \rangle \simeq \frac{k_B T}{m} t^2   + 
\langle ({\bf R}(0)- {\bf q}(0))^2  \rangle 
\left(  \delta(0)  + \frac{\hbar^2 t^2}{4 m^2 \delta^3(0)}   \right)
\ee
showing as expected a quadratic behavior with time.

\subsection{The Brownian or diffusion regime}

In this regime, as mentioned above, the thermodynamical equilibrium is already reached playing no role the details of the surface
such as the corrugation and the interaction potential ($\omega = 0$). This takes place 
at long times, that is, when $\eta t >> 1$. 
dynamics. After Doob's theorem \cite{doob}, the  classical VAF is now given by 
\begin{equation}\label{vaf-dr}
\mathcal{C}_{{\bf K}}(\tau) = \langle v_{{\bf K}}^2(0)  \rangle  e^{- \eta t} 
= \frac{k_B T}{m} e^{- \eta t} .
\end{equation}
which tell us that the corresponding correlation is decreasing exponentially with time.
The ISF and DSF in this classical framework are well known and given by \cite{salva8}
\begin{equation}
I({\bf K},t) = e^{- \chi^2  (e^{- \eta t} + \eta t - 1)}
\label{isf-isa-dr1}
\end{equation}
and 
\be
S({\bf K},\omega) = \frac{e^{\chi^2}}{\pi} \sum_{n=0}^{\infty} (-1)^n \frac{\chi^{2n}}{n!} 
\frac{(\chi^2 + n) \eta}{\omega^2+ \eta^2 (\chi^2+ n)^2}
\label{dsf1-dr1}
\ee
respectively, where the so-called shape parameter $\chi$ is defined by
\be\label{chi}
\chi = \frac{{\bf K}}{\eta} \sqrt{\langle v_{{\bf K}}^2(0)  \rangle}  =  {\bf K} \, \, {\bar l}
\ee 
which governs the dynamical coherence of the diffusion process. In this expression, ${\bar l}$ is the mean 
free path. It is well known that the time asymptotic behavior of the MSD gives the diffusion coefficient 
through Einstein's relation
\begin{eqnarray}\label{D}
D & = & \lim_{t \rightarrow \infty} \frac{1}{4 t} \langle |{\bf q}(t) - {\bf q}(0) |^2 \rangle  \nonumber \\
& = &   \lim_{t \rightarrow \infty} \int_0^t   dt' \langle v_{{\bf K}}(0)  v_{{\bf K}}(t') \rangle 
=  \frac{k_B T}{m \eta}    . 
\end{eqnarray}
The information about $D$ can also be extracted from the observable ISF and DSF. In this diffusion regime, we 
have that $\chi << 1$ and the ISF is given by a time exponential function 
\begin{equation}
I({\bf K},t) = e^{- {\bf K}^2 D t}
\label{isf-isa-dr2}
\end{equation}
and the DSF by a single Lorentzian function
\be
S({\bf K},\omega) \propto  \frac{{\bf K}^2 D}{\omega^2+ {\bf K}^4 D^2}
\label{dsf1-dr2}
\ee
which its FWHM is $\Gamma = 2 D {\bf K}^2$. Interestingly enough,
in the extreme opposite case, $\chi >> 1$, we approach the ballistic regime already discussed previously. 
In general, the continuous variation of the $\chi$-parameter can also be seen as a simple way to define 
the surface dynamical regime. When decreasing $\chi$, the corresponding DSF or line shape becomes narrower
and narrower. This gradual change of line shape is known as the {\it motinal narrowing effect} \cite{kubo}, 
going from a Gaussian to a Lorentzian line shape for the two extreme cases studied so far. 

In the quantum Langevin framework, it has been shown \cite{salva8} that the VAF is given by
\begin{equation}\label{Cv-q}
\mathcal{C}_{{\bf K}}(t) = \left( \frac{1}{m \beta} - i \frac{\hbar \eta}{2 m } \right) e^{- \eta t} -
\frac{2 \eta}{m \beta} \sum_{n=1}^{\infty} \frac{\nu_n e^{- \nu_n t} - \eta e^{- \eta t}}{\eta^2 - \nu_n^2}  .
\end{equation}
with $\nu_n = 2 \pi n /\hbar \beta$ (with $\beta = (k_B T)^{-1}$) being the so-called Matsubara frequencies. 
Quantum effects are important
at low temperatures, the long time behavior being mainly determined by the first term of the Matsubara series.
Thus, relaxation is no longer governed only by the damping constant. 
The ISF  in this quantum framework is then given by \cite{salva8}
\begin{equation}
I({\bf K},t) = e^{- \chi^2  (\eta t -\Phi(\eta t)) - i E_r t - {\bf K}^2 g(t)}
\label{isf-isa-dr-q}
\end{equation}
whith  
\begin{equation}
g(t)= \frac{2 }{m \beta} \sum_{n=1}^{\infty} \frac{\nu_n e^{- \eta t} - \eta e^{- \nu_n t} + \eta - \nu_n}
{\nu_n (\eta^2 - \nu_n^2)}
\end{equation}
and where it is clearly seen that the extra term ${\bf K}^2 g(t)$ in the argument of the ISF exponential 
is the difference with respect to the classical result. The DSF is now much more involved and can not be reduced 
to a simple analytical function. The diffusion coefficient is a complex number given by 
\begin{equation}\
D = \frac{k_B T}{m \eta} - i \frac{\hbar}{2m}
\end{equation}
whose real part is Einstein's law. The same result can be obtained from the MSD by considering only the
symmetric part of the autocorrealtion function. In any case, the limit to very small temperatures is questionable
since we are not taking into account the quantum noise correlation.

In the Bohmian framework, from Eqs. (\ref{vcor}) and (\ref{vaf-dr}), the VAF along the ${\bf K}$ 
direction is expressed as 
\begin{equation} \label{vcor-dr}
\mathcal{C}_{{\bf K}}(\tau) \equiv
\langle v_{{\bf K}}(0) \ \!
v_{{\bf K}}(\tau) \rangle = \frac{k_B T}{m} e^{- \eta t}
 + \langle (R(0)- q(0))^2  \rangle_{{\bf K}}  \frac{\dot {\delta}(0)}{\delta^2(0)} \dot{\delta}(t)
\end{equation}
and the dynamical equations governing this regime are given by Eqs. (\ref{leq}) and (\ref{pinney}).  
As mentioned before, the damped Pinney equation has not an analytical solution but it is possible to look for
an approximate one \cite{antonio2}. Eq. (\ref{pinney}) can be rewritten as
\be
\frac{d}{dt} \left( \frac{\dot{\delta}^2}{2} + \frac{\hbar^2}{8 m \delta^2} \right) =- \eta \dot{\delta}^2 \leq 0 .
\ee	
The expression inside brackets is essentially a positive definite quantity; the first term could be seen as the 
kinetic energy of the spreading and the second one as a potential function. At long times, due to the negative 
derivative (decreasing function with time), both terms tend to be negligible at different rates. In this regime,
the spreading acceleration is expected to be much smaller than the damping term $\eta \dot {\delta}$, leading
to a simple solution for Eq. (\ref{pinney}) to be
\be
\delta \sim \sqrt{\frac{\hbar}{m}} \eta^{-1/4} t^{1/4}     .
\ee
It is then straightforward to have that
\be
\frac{\ddot{\delta}}{\eta \dot{\delta}} \sim - \frac{3}{ 4\eta t}     ,
\ee
justifying the assumption made when $\eta t >> 1$.  Thus, Eq. (\ref{vcor-dr}) becomes
\begin{equation} \label{vcor-dr-1}
\mathcal{C}_{{\bf K}}(\tau) \equiv
\langle v_{{\bf K}}(0) \ \!
v_{{\bf K}}(\tau) \rangle \simeq \frac{k_B T}{m} e^{- \eta t} +  g_{{\bf K},0}   \, \, \eta^{-1/4}  t^{-3/4}
\end{equation}
with
\be
g_{{\bf K},0}= \frac{1}{4}  \sqrt{\frac{\hbar}{m}} \langle (R(0)- q(0))^2  \rangle_{{\bf K}}  \frac{\dot {\delta}(0)}{\delta^2(0)}
\ee
showing the time dependence of the Bohmian VAF. The time depedendent extra contribution goes 
with $t^{-3/4}$, typical from a dissipative behavior for the spreading of the Gaussian distribution 
function \cite{antonio2}.  The Bohmian stochastic trajectories, after Eq. (\ref{qtray}), are then expressed as
\be
{\bf R}(t) = {\bf q}(t) +  ({\bf R}(0) - {\bf q}(0)) \frac{1}{\delta(0)}   \sqrt{\frac{\hbar}{m}} 
\eta^{-1/4} t^{1/4}    .
\label{qtray-dr}
\ee

The ISF in this framework is then given by
\begin{equation}
I({\bf K},t) = e^{- \chi^2  (e^{- \eta t} + \eta t - 1)} e^{- {\bf K}^2 \frac{16}{5}   g_{ {\bf K},0}   \, \, \eta^{-1/4}  t^{5/4} }      .
\label{isf-isa-dr1}
\end{equation}
Now, at very long times, the argument of the first and second factors contributes linearly with $t$ and then 
\begin{equation}
I({\bf K},t) =  e^{- {\bf K}^2 (D + \alpha) t }      .
\label{isf-isa-dr2}
\end{equation}
with 
\be
\alpha = \frac{16}{5} g_{ {\bf K},0}  \eta^{-1/4} {\bar a}
\ee
where ${\bar a}$ is an average time value of the extremely slow varying function $t^{1/4}$.
The DSF or line shape is now expressed as 
\be
S({\bf K},\omega) \propto  \frac{{\bf K}^2 (D + \alpha)}{\omega^2+ {\bf K}^4 (D + \alpha)^2}
\label{dsf1-dr2}
\ee
which again a single Lorentzian function is obtained but with a higher FWHM given by 
$\Gamma = 2 (D + \alpha) {\bf K}^2$. The parameter $\alpha$ is zero at least when 
$\dot {\delta}(0) =0$.

In the diffusion regime, the corresponding MSD  is no longer linear with time 
\be
\langle {\bf R}^2 (t)  \rangle \simeq  2 D t + \langle ({\bf R}(0) - {\bf q}(0))^2 \rangle 
\frac{1}{\delta^2(0)}  \frac{\hbar}{m}
\eta^{-1/2} t^{1/2}
\ee
since the crossing term goes to zero at long times. This MSD also keeps the 
same dressing scheme of the stochastic trajectories, the first contribution is a particle contribution given in 
terms of the diffusion coefficient $D$ (behaving as in the classical case) and the second one comes from the 
wave spreading but with different time dependent behaviors. This dressing scheme for the MSD and its 
time dependence characterize the so-called Bohmian-Brownian motion \cite{salva7}. This slight deviation 
from the linearity could be seen as a weak anomalous diffusion process. \cite{raul}

\subsection{The intermediate regime}

In this intermediate regime, analytical results for ISF and DSF are only obtained if the classical VAF is assumed 
to  follow simple functional forms; in any case, numerical simulations of the Langevin equations have to be
carried out to extract the parameters involving the particular functional form assumed. 

In the classical framework, it is acceptable \cite{salva8} to assume that the VAF is well described by
\begin{equation}\label{vaf-ir}
\mathcal{C}_{{\bf K}}(\tau) = \langle v_{{\bf K}}^2(0)  \rangle = \frac{k_B T}{m} e^{- \eta t}
cos (\omega t + \delta) .
\end{equation}
where a temporary trapping of the adsorbate is expected to occur inside the wells of the corrugated 
surface interaction potential.  The $\omega$-parameter gives the frequency of these intrawell oscillations 
with a certain dephase $\delta$. Physically, this expression has the correct time behavior corresponding to 
the ballistic and Brownian regimes analyzed previously. As has been shown elsewhere, the ISF issue from
Eq. (\ref{vaf-ir}) has a more or less simple analytical expression leading to a DSF taking into account 
the intrawell motions which are of low energy quite close to the main quasielastic peak due to zero energy 
transfer. 

For massive particles, the mean interparticle distance is most of the time greater than the thermal de Broglie
wavelength $\lambda_B = \hbar / \sqrt{2 m k_B T}$ and quantum effects are only considered to be a correction
\cite{salva9}. The $I_2$ factor could be replaced by the classical Eq. (\ref{isf-isa2}) but this approximation 
is not good at small times. However, due to the fact the diffusion regime is reached at long times, the only 
quantum correction comes from the $I_1$ factor. Obviously, for light particles, where tunnelling can be present,
the approach is radically different.

The nice thing about the Bohmian framework with respect to the quantum one is that information of the
classical motion can still be used. Thus, the Bohmian VAF can now expressed as
\begin{equation} \label{vcor-ir}
\mathcal{C}_{{\bf K}}(\tau) \equiv
\langle v_{{\bf K}}(0) \ \!
v_{{\bf K}}(\tau) \rangle = \frac{k_B T}{m} e^{- \eta t} cos (\omega t + \delta)
+ \langle (R(0)- q(0))^2  \rangle_{{\bf K}}  \frac{\dot {\delta}(0)}{\delta^2(0)} \dot{\delta}(t)  .
\end{equation}
The adiabatic approximation could be still used in order to calculate the Bohmian stochastic trajectories. When the corrugation 
of the surface is strong enough this approximation is no longer valid and alternative solutions should be found
As mentioned before, the Pinney equation governing the width of the Gaussian density can not be solved 
analytically. A numerical solution has been obtained by Tsekov \cite{tsekov} and a first-order perturbation
solution by Haas et al. \cite{antonio2} where the acceleration term is assumed to be small, reproducing very
well the asymptotic behavior. For initially rapidly expanding wave packets, the damping term becomes 
dominant after a long period of time. With this in mind, the ISF and DSF are only known by numerical
calculations. Again, for an initial spreading velocity zero, the standard classical stochastic trajectories and 
VAF are recovered as well as the ISF and DSF.  

As a word of conclusion, in this work we have put in evidence that Bohmian stochastic trajectories are 
also able to describe surface diffusion processes when Ohmic friction, moderate surface temperatures and 
small coverages are assumed. An important difference can be seen when the initial spreading rate of the
Gaussian wave packet is considered zero or not. In particular, when this initial rate is not zero, the diffusion
process described in terms of Bohmian stochastic trajectories displays a weak anomalous diffusive behavior.  
Within this approach, the incoherent tunnelling regime  sould be carried out 
with success after our experience of applying it to the dissipative tunneling by a parabolic barrier 
\cite{vahid2}. It is true that the corresponding formalism should be extended to include the surface periodicity.
At the same time, this diffusion process could also be extended and described by scaled trajectories, recently
proposed to study dissipative dynamics \cite{vahid1,vahid2}  providing a smooth classical-quantum
transition.  When the Ohmic friction is not a good asumption then the generalized Langevin 
equation formalism is the appropriate dynamical equation together with colored noise \cite{pedro1}. These
interesting topics as well as to analyse the diffusion process in terms of continuous measurement \cite{salva7} 
are hopefully to be considered in the near future.

\vspace{2cm}
\noindent
{\bf Acknowledgements}
Support from the Ministerio de Ciencia, Innovaci\'on y Universidades under the 
Project FIS2017-83473-C2-1-P is acknowledged. I would  like to thank Prof. A. B. Nassar 
for helpful discussions in the Bohmian theory. 

\vspace{2cm}


%
\end{document}